# Evaluation of the Visibility and Artifacts of 11 Common Fiducial Markers for Image-Guided Stereotactic Body Radiation Therapy in the Abdomen

Jordan M. Slagowski, Ph.D.[1], Lauren E. Colbert, M.D., MSCR[2], Irina M. Cazacu, M.D.[3], Ben S. Singh, B.S.[3], Rachael Martin, Ph.D.[1], Eugene J. Koay, M.D., Ph.D.[2], Cullen M. Taniguchi, M.D., Ph.D.[2], Albert C. Koong, M.D., Ph.D.[2], Manoop S. Bhutani, M.D.[3], Joseph M. Herman, M.D., M.S.[2], and Sam Beddar, Ph.D.[1]

[1]Department of Radiation Physics, The University of Texas MD Anderson Cancer Center, Houston, TX 77030, USA

[2]Department of Radiation Oncology, The University of Texas MD Anderson Cancer Center, Houston, TX 77030, USA

[3]Department of Gastroenterology Hepatology and Nutrition, The University of Texas MD Anderson Cancer Center, Houston, TX 77030, USA

**Correspondence to:**

Sam Beddar, Ph.D.

Department of Radiation Physics, The University of Texas MD Anderson Cancer Center

1515 Holcombe Blvd, Unit 94, Houston, TX 77030, USA.

Tel: (713) 563-2609; Fax: (713) 563-2479; E-mail: abeddar@mdanderson.org

**Source of financial support / funding:** None

**Author responsible for statistical analysis:** Jordan M. Slagowski

**Correspondence to author responsible for statistical analysis:**

Jordan M. Slagowski, Ph.D.

Department of Radiation and Cellular Oncology, University of Chicago

5758 S Maryland Ave, MC 9006. Chicago, IL 60637, USA.

E-mail: jslagowski@uchicago.edu

**Abstract**

**Purpose:** The purpose of this study was to quantitatively evaluate the visibility and artifacts of commercially available fiducial markers in order to optimize their selection for image-guided stereotactic body radiation therapy (SBRT).

**Methods and Materials:** From six different vendors, we selected 11 fiducials commonly used in image-guided radiation therapy (IGRT); the fiducials varied in material composition (gold, platinum, carbon), shape (cylindrical, notched/linear, coiled, ball-like, step), and size measured in terms of diameter (0.28-1.0 mm) and length (3.0-20.0 mm). Each fiducial was centered in 4-mm bolus within a 13-cm-thick water-equivalent phantom. Fiducials were imaged with use of a simulation computed tomography (CT) scanner, a CT-on-rails system, and an onboard cone-beam CT system. Acquisition parameters were set according to clinical protocols. Visibility was assessed in terms of contrast (ΔHU) and the Michelson visibility metric. Artifacts were quantified in terms of relative standard deviation and relative streak artifacts level (rSAL). Twelve radiation oncologists ranked each fiducial in terms of clinical usefulness.

**Results:** Contrast and artifacts increased with fiducial size. For CT imaging, maximum contrast (2722 HU) and artifacts (rSAL = 2.69) occurred for the largest-diameter (0.75 mm) platinum fiducial. Minimum contrast (551 HU) and reduced artifacts (rSAL = 0.65) were observed for the smallest-diameter (0.28 mm) gold fiducial. Carbon produced the least severe artifacts (rSAL = 0.29). The survey indicated that physicians preferred gold fiducials with a 0.35- to 0.43-mm diameter, 5- to 10-mm length, and a coiled or cylindrical shape that balanced contrast and artifacts.

**Conclusions:** We evaluated 11 different fiducials in terms of visibility and artifacts. The results of this study may assist radiation oncologists who seek to maximize contrast, minimize artifacts, and/or balance contrast versus artifacts by fiducial selection.

**Introduction**

Stereotactic body radiation therapy (SBRT) involves delivering conformal high doses (≥5 Gy per fraction) of radiation over a reduced number of fractions (≤5) to body sites outside the central nervous system (1). SBRT is used for various cancer types in which ablative local doses are limited by nearby organs at risk (OARs) (2, 3). SBRT has demonstrated excellent local control and overall survival results in early-stage lung cancer, and its use is growing in popularity for other cancer types including prostate, pancreatic, renal, and adrenal cancers, in addition to oligometastatic disease (4-6). To achieve the high biologically effective target dose required for SBRT, treatment planning makes use of sharp dose falloff and tight planning target volume margins to minimize normal tissue toxicity. Volumetric imaging using cone-beam CT (CBCT), CT-on-rails (CTOR), or magnetic resonance imaging (MRI) prior to treatment is thus a necessity to visualize changes in soft tissue anatomy relative to planning target volumes.

Low-contrast lesions in the liver or pancreas are not easily visualized in CBCT images. To overcome this obstacle, high-contrast fiducials are often implanted into or nearby low-contrast gastrointestinal (GI) tumors under endoscopic ultrasound (EUS) guidance to serve as surrogates of tumor position (7, 8). During treatment setup, the fiducial markers help localize the tumor in the CBCT image. Even with CTOR, although OARs may be more easily visualized, intravenous contrast is not administered at the time of daily treatments, which may make isodense tumors impossible to visualize.

Many different types of fiducials of various shapes, sizes, and materials are available for implantation. The visualization of some fiducials can be more difficult than others due to differences in image artifacts and fiducial contrast (9, 10). Optimal fiducial selection requires consideration of the tradeoff between contrast and image artifacts, as well as clinical considerations such as the probability of fiducial migration and treatment modality (e.g., photons versus protons). Fiducial contrast is dominated by the x-ray energy spectrum and linear attenuation coefficient that is a function of the fiducial composition. Photoelectric absorption is the dominant photon interaction in the diagnostic energy range (keV) responsible for image contrast and varies as the cube of the atomic number of the attenuating medium. Conversely, high atomic number materials such as metal fiducials generate streak artifacts in x-ray CT images. Image artifacts may obscure low-contrast lesions, thereby increasing uncertainty during contour delineation or reducing setup accuracy before treatment delivery. Furthermore, artifacts in CT simulation images may introduce dosimetric errors during the treatment planning stage that are particularly concerning for proton therapy (11).

Currently, only a few studies have investigated the contrast and artifacts observed with CT imaging of the fiducials available for GI radiotherapy applications. *Chan et al.* (12) investigated six gold markers and one polymer marker and concluded that gold markers are visible in CT, CBCT, ultrasound, and 2D kV portal imaging, but the results were strictly qualitative. *Handsfield et al.* (13) performed a quantitative investigation of visibility and artifacts for gold, carbon, and polymer fiducials. Only fiducials from a single vendor and a cylindrical shape were considered. Platinum fiducials were also not considered, which may be of interest for clinics that use MRI-to-CT image fusion as part of the treatment planning process. *Nair et al.* (14) showed that the contrast of platinum fiducials implanted in liver tissue was significantly superior to gold seeds or coils when imaged with MRI. The purpose of this study was to provide a quantitative evaluation of fiducial visibility and artifacts to optimize fiducial selection for clinical applications, primarily in isodense GI cancers near OARs.

## Methods and Materials

*Fiducial markers*

Fiducial visibility and image artifacts were evaluated for 11 different fiducial markers selected from six vendors. Only fiducials suitable for implantation in the pancreas by EUS-guided placement were considered. The fiducials varied in terms of material composition (gold, platinum, carbon), shape (cylindrical, notched/linear, coiled, ball-like, step), and size measured in terms of diameter (0.28-1.0 mm) and length (3.0-20.0 mm). Carbon was considered since patients originally planned to receive proton radiation therapy (RT) may alternatively receive external photon beam RT after fiducial placement due to insurance or clinical considerations. Table 1 summarizes the relevant fiducial properties. The Gold Anchor fiducials, listed twice, have notches every 2 mm that allow the fiducials to be folded into a "ball-like" shape or placed as a "line" during implantation. The final configuration depends on tumor resistance and insertion technique.

**Table 1** Fiducial marker types and properties

| Manufacturer | Fiducial name | Material | Fiducial shape | Diameter (mm) | Length (mm) | Abbreviation |
|---|---|---|---|---|---|---|
| Carbon Medical Technologies | Acculoc Carbon Marker | Carbon | Cylindrical | 1.00 | 3 | Acculoc Carbon (C) |
| Naslund Medical AB | Gold Anchor | Gold | Linear | 0.28 | 10 | G.A. 10 mm linear (Au) |
| Naslund Medical AB | Gold Anchor | Gold | Ball | 0.28 | 10 | G.A. 10 mm ball (Au) |
| Naslund Medical AB | Gold Anchor | Gold | Linear | 0.28 | 20 | G.A. 20 mm linear (Au) |
| Naslund Medical AB | Gold Anchor | Gold | Ball | 0.28 | 20 | G.A. 20 mm ball (Au) |
| RadioMed Corporation/IBA | Visicoil | Gold | Coiled | 0.35 | 10 | Visicoil 10 mm (Au) |
| Cook Medical | EchoTip Ultra | Gold | Step | 0.43 | 5 | Cook Medical (Au) |
| Medtronic, Inc. | Beacon FNF needle | Gold | Cylindrical | 0.43 | 5 | Beacon FNF (Au) |
| Best Medical International, Inc. | Loose Gold Marker | Gold | Cylindrical | 0.80 | 3 | Best Medical (Au) |
| RadioMed Corporation/IBA | Visicoil MR | Platinum | Coiled | 0.35 | 10 | Visicoil 10 mm (Pt) |
| RadioMed Corporation/IBA | Visicoil MR | Platinum | Coiled | 0.75 | 5 | Visicoil 5 mm (Pt) |

*Phantom description*

Fiducials were placed on top of 2-mm Superflab bolus (Mick Radio Nuclear Instruments, Inc., Mt. Vernon, NY, USA) and 7-cm Virtual Water (Standard Imaging, Middleton, WI, USA) as shown in Figure 1. The fiducials were positioned perpendicular to the longitudinal axis of the couch which maximized the volume of each fiducial in a given image slice. An additional 2-mm bolus and 6-cm Virtual Water were placed on top of the fiducials before imaging. The 13.4 cm thick phantom measured 30 cm by 30 cm. The uniform phantom design facilitates the evaluation of artifacts since the images can be split into three regions consisting of fiducial, phantom background, and artifacts.

**Figure 1.** The uniform bolus and Virtual Water phantom are shown. Markings and in-room lasers were used to align the phantom with the imaging system isocenter (A). Two fiducials were placed 3 cm apart on top of a 2-mm-thick layer of bolus placed on top of 7 cm of Virtual Water phantom (B). An additional 2 mm of bolus and 6 cm of Virtual Water were placed on top of the fiducials (C).

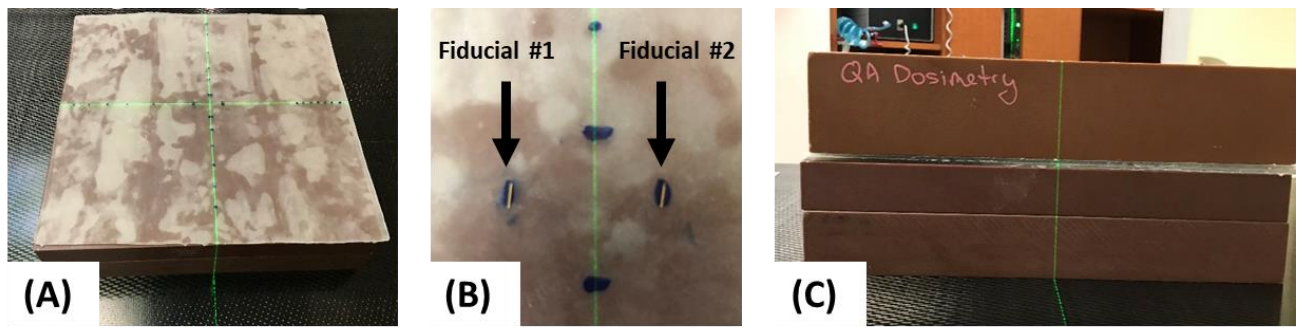

*Imaging modalities and acquisition parameters*

Each of the fiducials were placed within the phantom one at a time and imaged with a 16-slice CT simulation scanner (Brilliance Big Bore, Philips Medical Systems, Best, Netherlands), a CBCT system (TrueBeam OBI, Varian Medical Systems, Palo Alto, CA, USA), and a 16-slice CTOR scanner (GE LightSpeed RT16, GE Healthcare, Waukesha, WI, USA) (15). The phantom and fiducials were centered with respect to each imaging system's isocenter using in-room lasers and phantom markings to improve the reproducibility of set-up across each CT acquisition. Acquisition parameters were selected for each imaging modality based on institutional protocols for GI imaging, as summarized in Table 2. The CTOR exposure level may be adjusted from the default value of 84 mAs at the treatment machine, as needed. To investigate the impact of image noise on fiducial detectability a second exposure level of 250 mAs was also evaluated for the CTOR machine. The x-ray tube potential varied slightly, from 120 to 125 kV. Voxel dimensions were comparable across all imaging modalities. Slice thickness ranged from 2.0 mm for the CBCT system to 3.0 mm for the CT simulation scanner. Slice thickness is an important parameter that impacts the fiducial visibility due to volume averaging and image noise.

**Table 2** Image acquisition parameters

| Imaging system | Scanner type | kV | mAs | Slice thickness (mm) | Voxel dimensions (mm²) | Image bits |
|---|---|---|---|---|---|---|
| Philips Brilliance Big Bore | CT simulator | 120 | 300 | 3 | 0.98 x 0.98 | 12 |
| Varian TrueBeam OBI | CBCT | 125 | 1074 | 2 | 0.91 x 0.91 | 16 |
| GE LightSpeed RT16 | CT-on-rails | 120 | 84 | 2.5 | 0.98 x 0.98 | 12 |

*Fiducial visibility assessment*

Fiducial visibility was evaluated in terms of contrast ($\Delta HU$) and the Michelson visibility metric ($V$). Fiducial contrast was defined as:

$$\Delta HU = \overline{x_f} - \overline{x_b} \qquad (1)$$

Here, $\Delta HU$ represents the signal difference between the mean of image pixel values in a region-of-interest (ROI) located within the fiducial, $\overline{x_f}$, and a nearby background ROI, $\overline{x_b}$. The ROI, $x_f$, was selected as the 10 mm² area within the fiducial corresponding to the region of maximum intensity. The background ROI, $x_b$, was located 15 mm from the center-of-mass of the fiducial and perpendicular to the long axis of the fiducial in a 20 mm² region with minimal artifacts. For the CTOR machine, the contrast-to-noise ratio (CNR) was computed as CNR = $\Delta HU/\sigma_b$, where $\sigma_b$ denotes the noise standard deviation in the background ROI, to evaluate fiducial detectability.

The purpose of the Michelson visibility metric (equation 2) was to quantify contrast in terms of the brightest region of the fiducial, independent of fiducial or ROI size.

$$V = \frac{x_{max} - x_{min}}{x_{max} + x_{min}} \qquad (2)$$

Here, $x_{max}$ represents the maximum signal within a 30 by 30 mm² ROI defined about the center-of-mass of the fiducial, whereas $x_{min}$ represents the minimum value. Michelson visibility is bounded between 0 and 1 with a value of 1 corresponding to maximum visibility.

*Fiducial artifact-level assessment*

The presence of highly attenuating objects such as metal fiducials along x-ray paths traversing the object results in photon starvation. After image reconstruction, photon starvation manifests as streak artifacts and areas of signal void in regions adjacent to the implanted fiducial.

Streak artifacts were quantified using the relative streak artifacts level (rSAL) metric (16).

$$rSAL = \frac{|TV(x_{artifact}) - TV(x_{ref})|}{TV(x_{ref})} \qquad (3)$$

Here, $x_{artifact}$ denotes an artifact image, $x_{ref}$ denotes a reference image free of fiducial induced artifacts, and $TV(x)$ is the total variation function. The image $x_{artifact}$ was obtained by defining a 30 by 30 mm² square ROI about the center-of-mass of the fiducial and manually segmenting the fiducial from the square ROI yielding an image, $x_{artifact}$, containing only phantom background and artifacts. The reference image, $x_{ref}$, was obtained by imaging the phantom with no fiducials present. For consistency, the region corresponding to the fiducial in the image $x_{artifact}$ was also removed from $x_{ref}$. The total variation function, *TV(x),* was defined as:

$$TV(x) = \sum_{i,j} \sqrt{(x_{i+1,j} - x_{i,j})^2 + (x_{i,j+1} - x_{i,j})^2} \qquad (4)$$

Larger values of *TV(x)* indicate a higher level of streak artifacts.

The relative standard deviation (rStdDev), $\sigma_{ref}^{artifact}$, was used to further quantify the level of artifacts present.

$$\sigma_{ref}^{artifact} = \frac{\sigma_{artifact}}{\sigma_{ref}} \qquad (5)$$

Here, $\sigma_{artifact}$ and $\sigma_{ref}$ denote the standard deviations of image values in the artifact image, $x_{artifact}$, and reference image, $x_{ref}$, respectively. For fiducials producing no artifacts, the standard deviation of the artifact image, $\sigma_{artifact}$, is expected to be comparable to the standard deviation of the reference image, $\sigma_{ref}$, free of artifacts. As the artifact level increases, the standard deviation of the artifact image is expected to increase, resulting in a larger relative standard deviation. Thus, values of $\sigma_{ref}^{artifact}$ near 1.0 indicate that minimal artifacts are present, whereas larger values of $\sigma_{ref}^{artifact}$ indicate more severe artifacts.

*Survey of fiducial image quality*

Five faculty radiation oncologists and seven radiation oncology residents or fellows each rated the 11 different fiducial markers in terms of clinical usefulness. Participants were instructed to consider both the fiducial visibility and artifact level as seen on CT simulator and CBCT images of each marker placed in the water equivalent phantom. Each radiation oncologist was asked to rank each image (22 total) on a scale of 1 to 5 as follows: 1 = not useful at all for clinical practice, 3 = acceptable/sufficient for clinical practice, and 5 = ideal for clinical use. The differences between mean faculty and resident rankings were assessed by a right-tailed paired Wilcoxon signed-ranked test. Statistical significance was defined at the p-value < 0.05 threshold.

The Pearson correlation coefficient was also reported.

## Results

### *Fiducial visibility and artifacts*

Figure 2 presents images of each fiducial acquired with the CT simulator, CBCT, and CTOR systems. The images are organized in order of decreasing image contrast ($\Delta HU$) for the CT simulator. The bar plots present metrics of contrast and artifacts normalized to maximum values for the CT simulator images.

The maximum contrast value was observed for the 0.75-mm-diameter coiled platinum fiducial (Visicoil 5 mm) for each of the imaging systems. Figure 2 shows that the fiducial contrast and image artifact level increase as the diameter of the fiducial (Table 1) increases, as expected, except for the low atomic number carbon fiducial. The minimum contrast value was observed for the 0.28-mm-diameter linear-shaped gold fiducials. The contrast of the Gold Anchor fiducials was improved when the fiducial was deployed in the ball-like configuration. Likewise, a higher level of artifacts may be observed for the ball-like configuration. Qualitatively, fewer streak artifacts were present for the carbon and linear gold fiducials.

The qualitative observations were confirmed by the quantitative image quality metrics presented in Table 3. The mean signal difference between the Visicoil 5-mm fiducial and a nearby background region measured 2722 HU, 2824 HU, and 4798 HU for the CT simulator, CTOR, and CBCT systems, respectively, versus 551 HU, 889 HU, and 865 HU for the Gold Anchor 20-mm linear fiducial. The largest rSAL and relative standard deviation were observed for the Visicoil 5-mm platinum and Best Medical gold fiducials across all imaging modalities. The least severe artifacts were observed, as expected, for the carbon fiducial.

**Figure 2.** Fiducial markers within a uniform phantom and imaged with a CT simulator, CTOR, and CBCT are shown. The bar plots show the contrast and relative streak artifacts level (rSAL) metrics normalized to maximum contrast and rSAL for the CT simulator results. The window range is [-500, 1000] HU. Abbreviations for the fiducial labels are defined in Table 1.

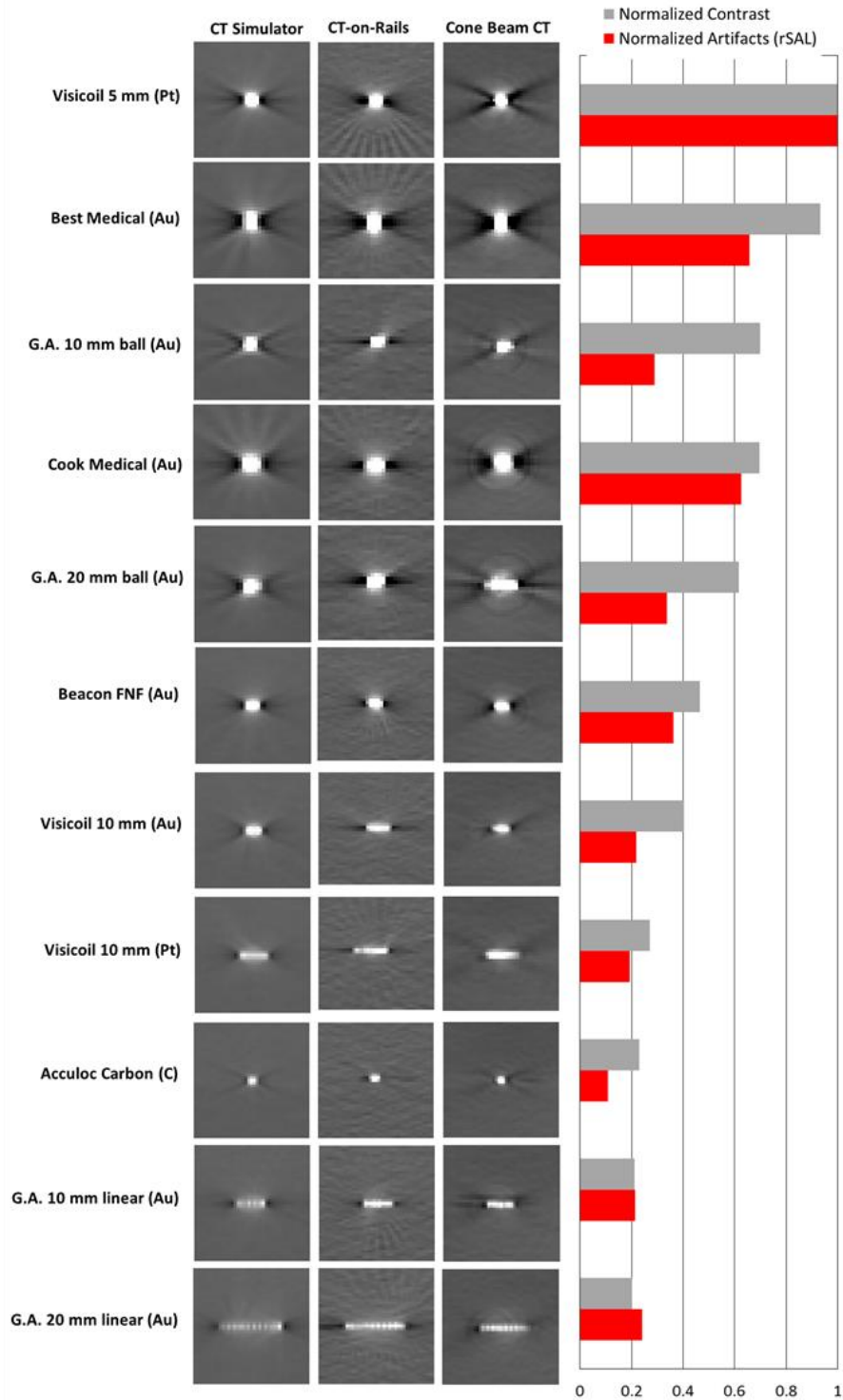

Fiducial rankings for each metric are displayed in parentheses in Table 3. A ranking of 1 of 11 indicates maximum contrast or minimal artifacts. Conversely, a ranking of 11 indicates inferior contrast or a high artifact level. A comparison of rankings will be helpful in selecting a fiducial that balances the tradeoff between contrast and artifacts. Of note, the Gold Anchor 10-mm gold fiducial in a ball-like configuration had an average contrast value of 4.5 of 11 and an average artifact level of 5.0 of 11, averaged across the three imaging modalities. This suggests that the fiducial design achieved a balance between contrast and artifacts with average metrics for both ranking in the top half of the fiducials evaluated.

Figure 3 presents the CNR of each fiducial imaged as a function of tube exposure for CTOR. The CNR increased as the exposure increased from 84 mAs to 250 mAs, as expected, for each fiducial.

**Figure 3.** The contrast-to-noise ratio is plotted for imaging performed with CT-on-rails at 84 mAs and 250 mAs for 11 different fiducial configurations. The error bars plotted for the Beacon FNF fiducial represent ±1 standard deviation uncertainty estimated from a series of 10 independent measurements and phantom set-ups. Abbreviations for the fiducial labels shown in the legend are defined in Table 1.

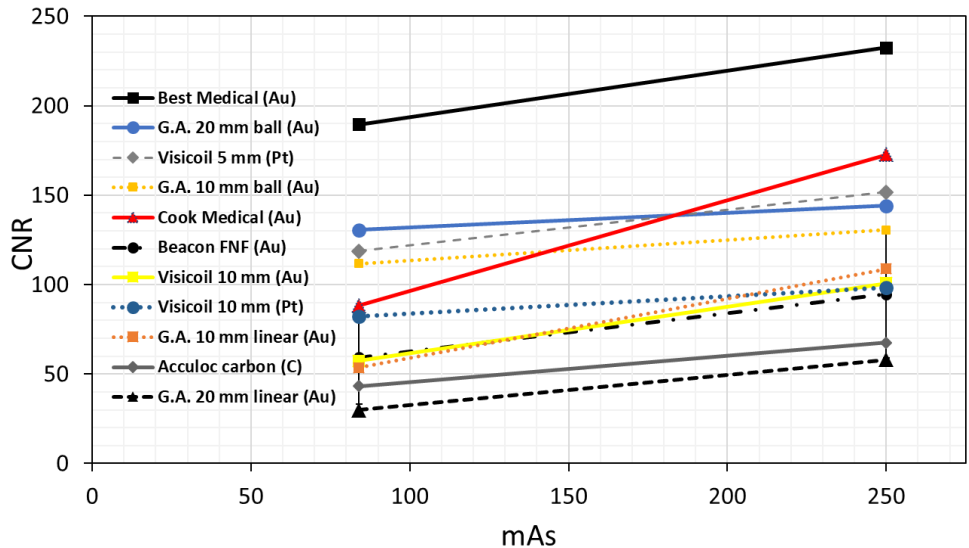

**Table 3** CT simulator, CT-on-rails, and cone beam CT fiducial contrast and artifacts

| | CT Simulator | | | | CT-on-rails | | | | Cone-beam CT | | | |
|---|---|---|---|---|---|---|---|---|---|---|---|---|
| | Contrast | | Artifacts | | Contrast | | Artifacts | | Contrast | | Artifacts | |
| Fiducial | ΔHU | Michelson | rStdDev | rSAL | ΔHU | Michelson | rStdDev | rSAL | ΔHU | Michelson | rStdDev | rSAL |
| Visicoil 5 mm (Pt) | 2722 (1) | 1.00 (1) | 15.6 (11) | 2.69 (11) | 2824 (1) | 1.00 (1) | 5.6 (10) | 1.68 (11) | 4798 (1) | 1.00 (1) | 7.5 (10) | 2.69 (11) |
| Best Medical (Au) | 2539 (2) | 0.94 (2) | 11.5 (10) | 1.77 (10) | 2753 (2) | 1.00 (1) | 5.8 (11) | 1.62 (10) | 4723 (2) | 1.00 (1) | 7.8 (11) | 2.42 (10) |
| G.A. 10 mm ball (Au) | 1903 (3) | 0.62 (5) | 5.5 (6) | 0.78 (6) | 2194 (5) | 0.72 (5) | 2.4 (4) | 0.47 (4) | 2270 (5) | 0.76 (5) | 2.7 (6) | 0.73 (6) |
| Cook Medical (Au) | 1898 (4) | 0.78 (3) | 9.7 (9) | 1.68 (9) | 2300 (4) | 1.00 (1) | 4.6 (9) | 1.21 (9) | 2873 (4) | 1.00 (1) | 6.5 (9) | 1.95 (9) |
| G.A. 20 mm ball (Au) | 1678 (5) | 0.65 (4) | 6.5 (7) | 0.91 (7) | 2621 (3) | 1.00 (1) | 4.4 (8) | 1.08 (8) | 2964 (3) | 0.84 (4) | 4.5 (8) | 1.58 (8) |
| Beacon FNF (Au) | 1266 (6) | 0.52 (6) | 6.8 (8) | 0.97 (8) | 1527 (6) | 0.72 (6) | 2.8 (7) | 0.64 (6) | 1710 (6) | 0.74 (6) | 3.5 (7) | 0.92 (7) |
| Visicoil 10 mm (Au) | 1096 (7) | 0.36 (7) | 4.3 (5) | 0.59 (4) | 1133 (8) | 0.38 (10) | 2.4 (5) | 0.51 (5) | 1379 (7) | 0.48 (8) | 2.4 (4) | 0.51 (5) |
| Visicoil 10 mm (Pt) | 739 (8) | 0.08 (11) | 3.5 (3) | 0.52 (2) | 1253 (7) | 0.45 (7) | 1.9 (3) | 0.46 (3) | 1324 (8) | 0.37 (9) | 2.0 (2) | 0.40 (2) |
| Acculoc Carbon (C) | 628 (9) | 0.15 (9) | 3.1 (1) | 0.29 (1) | 923 (10) | 0.42 (8) | 1.4 (1) | 0.20 (1) | 998 (9) | 0.50 (7) | 1.6 (1) | 0.33 (1) |
| G.A. 10 mm linear (Au) | 572 (10) | 0.11 (10) | 3.9 (4) | 0.57 (3) | 1071 (9) | 0.40 (9) | 1.9 (2) | 0.35 (2) | 946 (10) | 0.24 (10) | 2.3 (3) | 0.49 (4) |
| G.A. 20 mm linear (Au) | 551 (11) | 0.17 (8) | 3.3 (2) | 0.65 (5) | 889 (11) | 0.30 (11) | 2.6 (6) | 0.68 (7) | 865 (11) | 0.20 (11) | 2.6 (5) | 0.48 (3) |

*Abbreviations:* rSAL = relative streak artifacts level, rStdDev = relative standard deviation. Fiducial abbreviations are defined in Figure 1.

*Survey of fiducial image quality*

Table 4 presents the results from the physician survey of fiducial image quality. The Beacon FNF and Visicoil 10-mm gold markers were the two highest-ranked fiducials for both CT and CBCT imaging (mean ranking, 4.1 and 4.2, respectively). The Gold Anchor 20-mm fiducial marker (linear configuration, mean ranking, 2.1) was ranked the lowest for CT imaging, and the Best Medical fiducial (mean ranking, 2.4) was ranked the lowest for CBCT. Physicians preferred the Gold Anchor fiducials in the ball-like configuration versus the linear configuration. No significant difference was observed between faculty and resident rankings (p-value = 0.36). The correlation between mean faculty and resident rankings was 0.76.

**Table 4** Physician survey of fiducial marker image quality results

| Fiducial | CT-Sim $\mu \pm \sigma$ | CT-Sim Ranking | CBCT $\mu \pm \sigma$ | CBCT Ranking |
|---|---|---|---|---|
| Beacon FNF (Au) | 4.3 $\pm$ 0.8 | 1 | 3.8 $\pm$ 0.8 | 2 |
| Visicoil 10 mm (Au) | 4.3 $\pm$ 0.9 | 2 | 4.0 $\pm$ 1.0 | 1 |
| Visicoil 5 mm (Pt) | 3.5 $\pm$ 1.0 | 3 | 2.9 $\pm$ 1.2 | 7 |
| G.A. 20 mm ball (Au) | 3.3 $\pm$ 1.1 | 4 | 2.6 $\pm$ 1.1 | 8 |
| G.A. 10 mm ball (Au) | 3.0 $\pm$ 0.9 | 5 | 3.1 $\pm$ 0.9 | 5 |
| Cook Medical (Au) | 2.9 $\pm$ 1.3 | 6 | 2.6 $\pm$ 1.7 | 8 |
| Visicoil 10 mm (Pt) | 2.9 $\pm$ 0.8 | 6 | 3.3 $\pm$ 1.0 | 4 |
| Best Medical (Au) | 2.8 $\pm$ 1.2 | 8 | 2.4 $\pm$ 1.3 | 11 |
| Acculoc Carbon (C) | 2.8 $\pm$ 1.3 | 8 | 3.4 $\pm$ 1.2 | 3 |
| G.A. 10 mm linear (Au) | 2.4 $\pm$ 1.1 | 10 | 2.9 $\pm$ 1.1 | 6 |
| G.A. 20 mm linear (Au) | 2.1 $\pm$ 0.9 | 11 | 2.6 $\pm$ 1.0 | 8 |

*Abbreviation:* CBCT = cone-beam CT. Fiducial abbreviations are defined in Figure 1.

**Discussion**

This study evaluated the visibility and artifacts of 11 commercially available fiducial markers from six different vendors across three different volumetric imaging systems for application to image-guided SBRT. The slight difference in atomic number between platinum and gold was shown to be less important than fiducial size and other effects such as volume averaging. Carbon fiducials presented an alternative option to gold and platinum fiducials, yielding minimal artifacts and moderate contrast. An additional benefit of carbon is the reduced dosimetric uncertainty relative to gold or platinum making it the preferred fiducial material for proton therapy. This work, however, only considered CT and CBCT imaging. Carbon contrast is expected to be lower in 2D kV or MV portal images. Additionally, the minimum detectable contrast level of carbon markers should be determined in the presence of CBCT streak artifacts.

      Fiducial marker selection for IGRT in the abdomen requires careful consideration of the tradeoff between contrast and artifacts as well as awareness of dosimetric uncertainty and treatment modality. Image artifacts in CT simulation images may adversely affect the delineation of target structures and OARs. Insufficient contrast in CBCT images acquired before treatment delivery may increase set-up uncertainty and IGRT accuracy. The quantitative results presented in this study will aid the radiation oncologist in optimizing fiducial selection for IGRT. If high contrast is prioritized over reduced image artifacts, the results of this study show that one should choose the Visicoil 5-mm platinum, Best Medical gold, Cook Medical gold, or Gold Anchor (ball-like configuration) fiducials. Conversely, if one favors reduced image artifacts in CT simulation images and is willing to accept reduced contrast during IGRT, one ought to choose the Gold Anchor (linear configuration), Acculoc carbon, or Visicoil-10 mm platinum fiducials. Finally, if one prefers to balance contrast and image artifacts, the Beacon FNF gold or Visicoil

10-mm gold fiducials are the best option.

A survey of 12 radiation oncologists ranked the Beacon FNF gold and Visicoil 10-mm gold fiducials highest in terms of clinical usefulness. The Beacon FNF and Visicoil gold fiducials were the only fiducials to receive average rankings exceeding 4.0 for CT imaging. Of note, the two fiducials preferred in the physician survey were those that balanced contrast and artifacts as indicated by the quantitative metrics. The two fiducials receiving the lowest ranking for CT imaging were the Gold Anchor 10-mm ($2.4 \pm 1.1$) and 20-mm ($2.1 \pm 0.9$) fiducials deployed in a linear configuration. The low physician ranking is validated by the quantitative metrics (Table 3) which showed these fiducials demonstrated the lowest image contrast (ΔHU).

In addition to image quality, certain practical aspects must be considered during the fiducial selection process. Fiducial selection must include a discussion of the placement-related technical challenges, complications, and migration probability unique to each fiducial type. Fiducials can be delivered using 19-, 22-, or 25-gauge needles or multifiducial delivery systems. The size and shape of the fiducial marker typically dictate the type of needle to be used. Fiducials with a broad diameter (>0.75 mm) require a 19-gauge needle for deployment. Studies have shown that the use of this needle is feasible and safe; however, its stiffness can compromise fiducial placement, especially in challenging anatomic locations or after surgery. In these situations, the 19-gauge needle can be replaced with a 22-gauge needle preloaded/backloaded with the adequate fiducial marker.

The probability of fiducial migration must also be considered. Fiducial migration rates have been shown to be low; nevertheless, this is an important consideration since migration may make the fiducial unusable for IGRT, decrease target coverage, or result in increased dose to normal tissue structures (7, 17). A study by *Khashab et al.* (17) demonstrated a 6.5% fiducial migration rate for traditional fiducials versus 3.8% for coiled fiducials; however, the difference was not statistically significant.

Finally, one must consider the patient's intended treatment modality and likelihood of transfer to a different modality during the fiducial selection step. For example, a patient originally selected for proton therapy or MR-guided RT may later be transferred to photon-based external beam RT. In these instances, one must consider fiducial contrast, artifacts, and dosimetric uncertainty for each of the different treatment modalities.

Several limitations of this work must be acknowledged. First, fiducials were imaged within a simple phantom consisting of only bolus and Virtual Water. In a clinical scenario, fiducials may be placed near regions containing air or bones that could influence the results of this study. Second, object motion was not considered since it was assumed that most GI cancers may be treated under the condition of breathhold to minimize respiratory motion. Cone beam CT streak artifacts resulting from heterogeneities or motion may reduce fiducial contrast. Future studies are required to evaluate the impact of CBCT image artifacts on IGRT accuracy. Third, image acquisition parameters were selected based on current clinical protocols. The impact of tube potential and exposure time on contrast-to-noise were described previously by *Hansfield et al* (13). In addition to tube parameters, the impacts of image reconstruction parameters, including pixel size, algorithm, and metal artifact correction, could be investigated as follow-up work. *Brook et al.* (18) demonstrated that metal artifact reduction software may improve tumor visibility near gold fiducial markers. However, *Brook et al.* focused on spectral CT, which is not readily available for RT simulation. Fourth, only a single fiducial orientation was investigated. All fiducials were positioned orthogonal (90°) to the longitudinal axis of the couch in order to maximize image contrast. To demonstrate the impact of fiducial orientation, the 5-mm platinum fiducial was imaged at different angular orientations corresponding to 0°, 45°, and 90° relative to the long axis of the treatment couch. Figure 4 shows that the Michelson contrast and rSAL increased as the fiducial rotated from parallel (0°) to perpendicular (90°) to the imaging volume's slice dimension and long axis of the treatment couch. The increase is expected as a greater portion of the fiducial is contained within an individual image slice compared with the parallel orientation, which increases photon absorption.

**Figure 4.** Fiducial relative streak artifacts level (rSAL) and contrast (Michelson) are plotted as a function of fiducial orientation. Orientation is defined with 0°, indicating that the long axis of the fiducial is parallel to the treatment couch and slice dimension of the image volume.

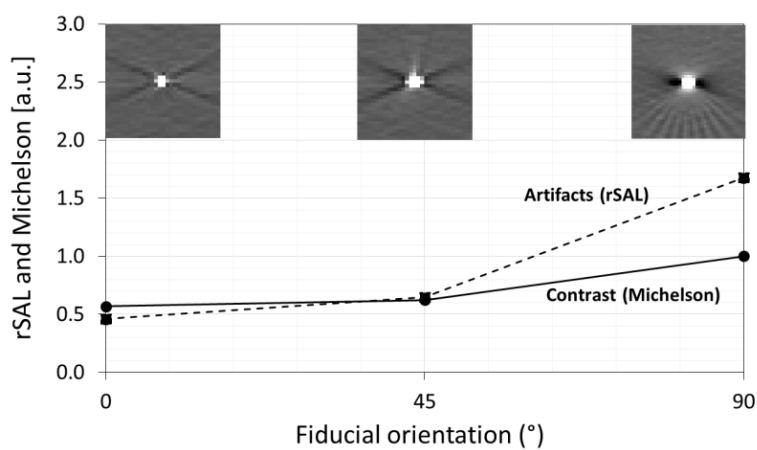

**Conclusion**

We evaluated 11 commonly used fiducials for image-guided SBRT in terms of visibility and artifacts under imaging performed with a CT simulator, CTOR, and on-board CBCT. The results will provide useful information when selecting a fiducial type with a preference towards high fiducial contrast, low image artifacts, or a balanced tradeoff between contrast and artifacts. A survey of radiation oncologists indicated that fiducials balancing image contrast and artifacts are preferred.

...end21